# PDMS microfluidic film for in vitro engineering of mesoscale neuronal networks

Taiki Takemuro,[a] Hideaki Yamamoto,[a] Shigeo Sato[a] and Ayumi Hirano-Iwata[a,b]

[a] *Research Institute of Electrical Communication, Tohoku University, Sendai 980-8577, Japan.*
[b] *WPI-Advanced Institute of Materials Research (WPI-AIMR), Tohoku University, Sendai 980-8577, Japan.*

**Abstract**

Polydimethylsiloxane (PDMS) microfluidic devices have become a standard tool for engineering cells and multicellular networks in vitro. However, the reservoirs, or through-holes where cells access the devices, are usually fabricated manually using a biopsy punch, making it difficult to create a large-scale array of small (<1 mm) reservoirs. Here, we present a fabrication process for a thin-film microfluidic device, or a microfluidic film (μFF), containing an array of through-holes. Holes as small as 100 μm by 100 μm spanning 10 mm by 10 mm are characterized. The geometry of the through-holes was precisely defined by the photoresist mould. A challenge in using the μFF for cell culture was air-bubble entrapments in the through-holes, which became more prominent with smaller holes. We show that this issue can be overcome using ethanol-mediated wetting of the PDMS surface, and demonstrate functional recording of cultured neuronal networks grown in μFFs. This technology opens new application of microfluidic devices to mesoscale systems comprised of several tens to hundreds of cells.

## 1. Introduction

Driven by concepts such as the organ-on-a-chip for pharmaceutical sciences, microfabrication technology continues to provide novel applications in cell engineering research.[1–3] Neuroscience was among the first fields to adopt cell culture on engineered scaffolds.[4–8] Since then, methods including photolithography, microcontact printing, and microfluidic devices have been utilized to provide novel *in vitro* platforms for fundamental and applied neuroscience studies.

Among the multiple approaches that can be used for cell patterning, polydimethylsiloxane (PDMS) microfluidic devices have now been accepted as a highly reproducible and stable platform.[9,10] For neuroscience applications, microfluidic devices have been used to confine distinct populations of neurons in separate chambers,[11,12] permit the growth of axons in defined orientations,[13–15] and record signal propagation within axons.[16–18] However, a limitation of the microfluidics approach is that the reservoirs, or the through-holes via which the cells enter the device, are generally fabricated separately from the microfabrication process, e.g., by manually creating mm-sized holes with a biopsy punch.[9,14] This becomes a major issue when designing neuronal networks with a restricted number of neurons, in the range of several tens to hundreds of cells. Although PDMS microstructures with well-defined through-holes have previously been fabricated using plasma etching,[19] capillary filling,[20,21] manual air blowing,[22] and gel spreading in open air,[23] application of the methods to the fabrication of cell engineering devices has rarely been reported.

Here, the details of a microfluidic film (μFF), a thin microfluidic device with μm-sized reservoirs, are reported for application in patterning mesoscale neuronal networks. A simple drop-casting of PDMS gel is shown to be sufficient for structuring the PDMS from a master mould prepared using SU-8 photoresists. However, air bubbles were easily trapped in the small reservoirs, thereby presenting a major challenge for this novel setup. An effective method to address this issue using ethanol and serum-containing medium to wet the PDMS surface and reduce its hydrophobicity is detailed. Finally, we demonstrate an application of the newly prepared device to pattern modular networks of primary neurons.

## 2. Materials and methods

### Device fabrication

The fabrication of μFF involves two steps: fabrication of a master mould and subsequent PDMS structuring. The master mould was fabricated by patterning two photoresist layers on a silicon substrate. A diced silicon (approximately 25×25 mm$^2$) was first cleaned in a piranha solution (a 1:1 mixture of concentrated $H_2SO_4$ and 30% $H_2O_2$) for 10 min, and the surface oxide was subsequently stripped in 5% HF for 5 s. Afterwards, a SU-8 3010 photoresist (Kayaku

Advanced Materials) was spin coated at 3000 rpm and subsequently baked for 1 and 5 min at 65 and 95 °C, respectively. Next, photolithography was performed using a mask aligner (Suss MBJ-4) and a chromium photomask with both reservoir and microchannel patterns. After a post-exposure bake for 1 and 3 min at 65 and 95 °C, respectively, the pattern was developed in a SU-8 Developer and rinsed twice in 2-propanol.

The abovementioned procedure was repeated using a thicker photoresist to form the second layer. For this step, the photoresist was replaced with SU-8 3050 (Kayaku Advanced Materials), which was spin coated at 1500 rpm, and a photomask with only the reservoir patterns was used. A pre-exposure bake was performed for 1 and 20 min at 65 and 95 °C, respectively, followed by a post-exposure bake for 1 and 5 min at 65 and 95 °C, respectively. Details regarding the pattern geometry are provided in Section 3. Individual patterns were arranged in a matrix of 11×11 or 12×12 to fit in a square region of 10×10 mm$^2$.

Using the fabricated SU-8 mould, the μFF was produced using Sylgard 184 (Dow Corning). The base and curing agents were mixed in a 10:1 ratio and degassed. Subsequently, 6 μL of the mixed gel was drop-casted to the edge of the 10×10 mm$^2$ square using a P20 micropipette. The gel was then allowed to spread for 10–30 min. Finally, PDMS was thermally cured in an oven for 3 h at 60 °C and peeled off from the mould substrate using Dumont #5 forceps.

**Sample characterization**

Three-dimensional topography of the photoresist and PDMS microfluidic device was analyzed using confocal microscopy (Keyence VK-X260). All the samples were imaged without prior surface coating. The data were analyzed using MultiFileAnalyzer software (Keyence).

Bubble trapping in reservoirs was quantified under a stereo microscope. The μFF was first sterilized under UV light for 30 min and attached to a glass coverslip (Matsunami C018001; diameter, 18 mm; thickness 0.17 mm) coated with poly-D-lysine (PDL; Sigma P-0899). The sample was subsequently immersed in a minimum essential medium (MEM; Gibco 11095-080) and imaged using a stereo microscope and digital camera. The use of the phenol red-containing MEM, rather than water, substantially eased the observation of bubbles.

**Cell culture**

Prior to cell culture, the μFF attached to a PDL-coated coverslip was immersed in a neuronal plating medium [MEM (Gibco 11095-080) + 5% fetal bovine serum + 0.6% D-glucose] and stored overnight in a cell-culture incubator. Rat cortical neurons were obtained from the cortices of E18 embryos and cultured using previously published protocols.[24] All procedures were approved by the Tohoku University Center for Laboratory Animal Research, Tohoku University

(approval number: 2017AmA-001-1) and the Tohoku University Center for Gene Research (2019AmLMO-001). Briefly, the dissociated cells were suspended in the plating medium and plated at a concentration of 3.8–5.0×10$^4$ cells/cm$^2$. The coverslip with the µFF was then transferred and placed upside down in a cell-culture dish with glial cells growing in the N2 medium. At 4 days in vitro (DIV), the cultured cells were transfected with the fluorescent calcium probe GCaMP6s (Addgene viral prep #100843-AAV9)[25] using adeno-associated virus vectors. Spontaneous and evoked network activities were recorded at 10 DIV via fluorescence calcium imaging.[24,26,27] Statistical analyses of neural correlations were restricted to networks bearing 70−120 neurons to focus on the impact of network morphology. NeuO dye (STEMCELL Technologies #01801)[28] was used to fluorescently stain live neurons by incubating the cells with 0.2 µM NeuO for 30 min.

## 3. Results and discussion
### Fabrication of the PDMS microfluidic film

The fabrication process of the µFF is schematically outlined in Fig. 1. The master mould was created by patterning two photoresist layers on a clean silicon substrate. SU-8 3010 was used to form the first layer and was patterned to create the reservoirs and microchannels (Fig. 1a and b). Subsequently, a thicker photoresist, SU-8 3050, was used for the second layer and patterned to create only the reservoirs (Fig. 1c and d). The PDMS prepolymer was then drop-casted to the edge of the patterned photoresist (Fig. 1e), thermally cured, and peeled off to release the completed μFF (Fig. 1f and g). Further details of the process are described in Section 2.

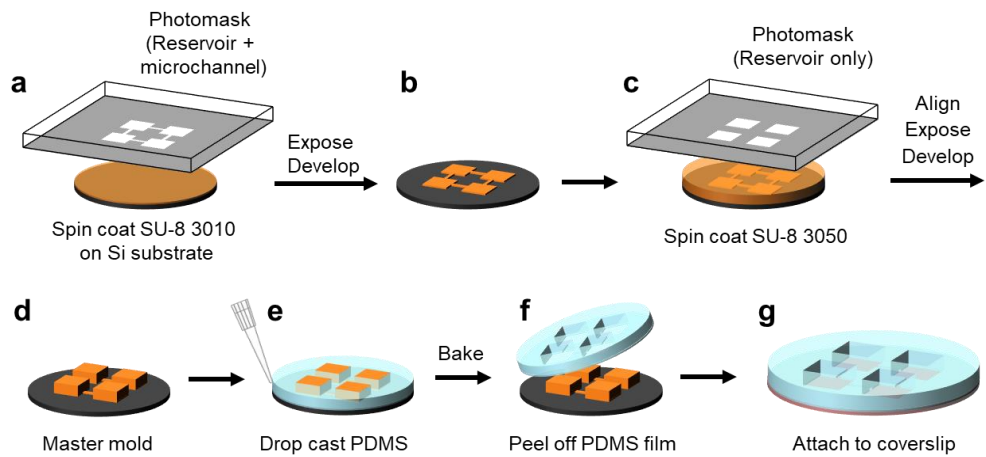

**Fig. 1 Fabrication of the µFF device.** (a, b) Patterning of a thin photoresist layer (approximately 10 µm high) to fabricate the reservoir and microchannel structures. (c, d) Patterning of a thick photoresist layer (approximately 90 µm high) to heighten the reservoir areas, completing the master mould fabrication. (e, f) Structuring and detaching the PDMS. (g) Attachment of the PDMS microfluidic device to a coverslip for cell culture.

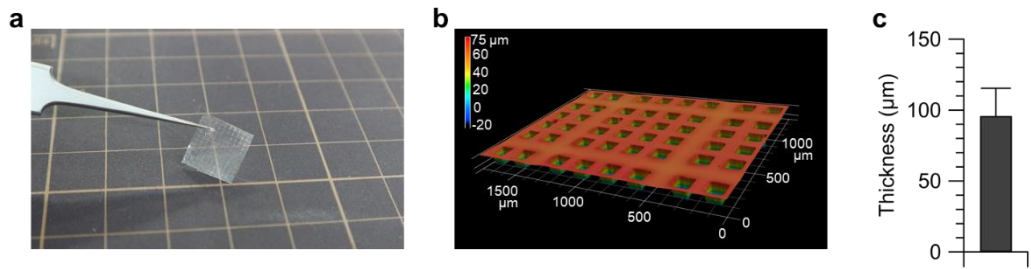

**Fig. 2 Characterization of the μFF device.** (a) Photograph of the μFF. (b) 3D confocal image of the μFF with a 16-module micropattern from the top. (c) Mean thickness of the PDMS film analyzed using laser confocal microscopy. Error bar, S.D.

The μFF fabricated herein was 96.0 ± 19.4 μm (mean ± S.D.; $n$ = 15 observations from 5 samples) thick and designed to be 10 mm by 10 mm in size (Fig. 2). In conventional microfluidic devices, the reservoirs are created by manually punching mm-sized holes, while those of the μFF were defined by the second-layer photoresist of the master mould, enabling precise definition of an array of reservoirs as small as 100×100 μm$^2$ (Fig. 2b). The minimum feature size of the reservoirs was restricted by the area necessary for proper growth of neurons, rather than inherent limitations of the lithography process.

The micropatterns consisted of reservoirs for cell adhesion (areas: 100×100, 200×200, or 400×400 μm$^2$) and microchannels for neurite growth (widths: 2, 5, or 10 μm). Schematic illustrations of representative micropatterns are presented in Fig. 3a–c along with confocal micrographs of the prepared μFF. Dissociated neurons cultured in these structures developed to form a network with modular organization, a network connectivity which is evolutionarily conserved in the nervous systems of animals.[29] From the confocal imaging, the height of microchannels was estimated as 9.2 ± 0.3 μm ($n$ = 10 observations from 2 samples; Fig. 3d–f). The widths of the channels were 3.6 ± 1.4, 5.5 ± 0.4, and 11.3 ± 0.3 μm (each $n$ = 10 observations from 2 samples) for the microchannels patterned using photomasks of 2, 5, and 10 μm, respectively (Fig. 3g). Manipulation of channel widths has been demonstrated to be effective for controlling the number of axons that enter the channels,[13] which is a critical parameter that defines the degree of modularity in mesoscale networks.[30]

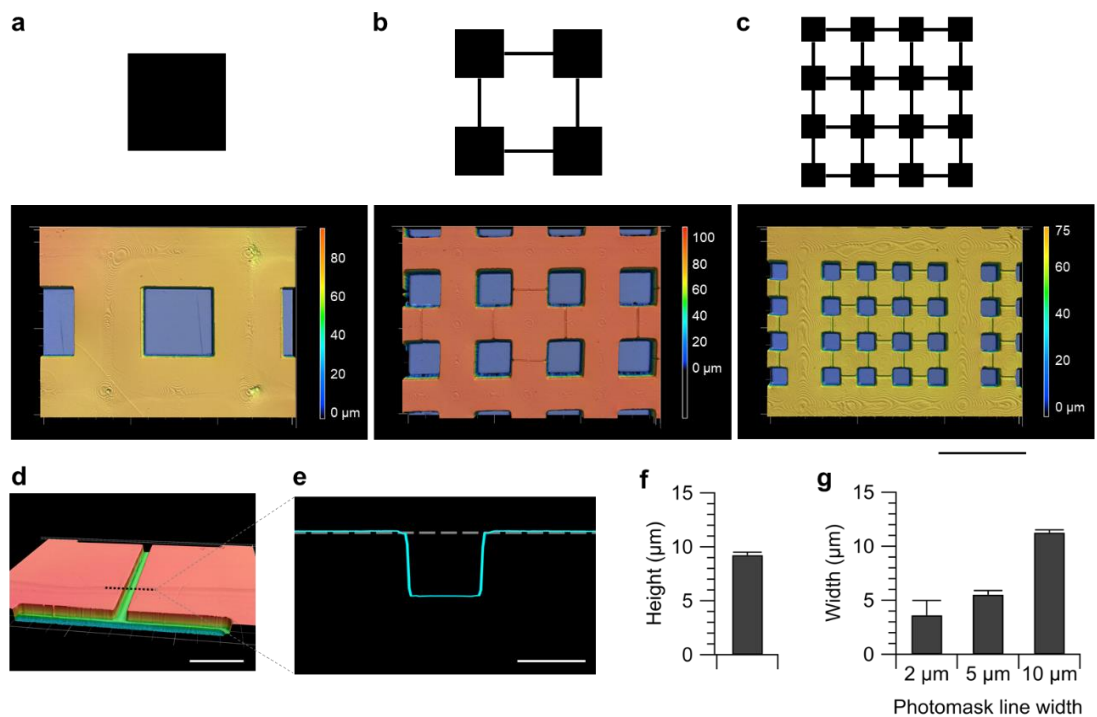

**Fig. 3 Characterization of the microstructures.** (a, b, c) Schematic illustration (top) and 3D confocal micrographs of the PDMS µFF with 1- (a), 4- (b), and 16-module (c) micropatterns. All imaging was performed from the side of the film with the microchannels. (d) High magnification observation of the microchannel region in a 4-module micropattern with 10 µm-wide channels. (e) Line profile of the section marked in (d). (f, g) Mean height and width of the microchannels. Error bars, S.D. Scale bars, 500 µm (a–c), 50 µm (d), and 10 µm (e).

**Neuronal cultures in the microfluidic film**

After attaching the µFF to a PDL-coated coverslip, the sample was immersed in the neuronal plating medium overnight to prepare for cell seeding. A major issue that arose at this stage was gas entrapment in the reservoirs (Fig. 4a, left), which became increasingly problematic as the reservoir sizes were reduced (Fig. 4b).

To prevent bubble trapping, the sample was soaked in 99.5% ethanol prior to immersion in the neuronal plating medium.[31,32] This procedure was highly effective for reducing the number of air bubbles trapped in the reservoirs and microchannels (Fig. 4a, right). For instance, the fraction of reservoirs with air bubbles was reduced from 99.9% to 6.1% for reservoirs with the size of 100×100 µm$^2$. This bubble reduction effect was prominent in other reservoir sizes as well (Fig. 4b). Prior to cell plating, the entire medium was exchanged with a fresh medium to eliminate residual ethanol. At this stage, hardly any bubbles remained, possibly due to the elevated temperature during the storage of the sample in the $CO_2$ incubator and wetting of the surface by the serum proteins contained in the plating medium.

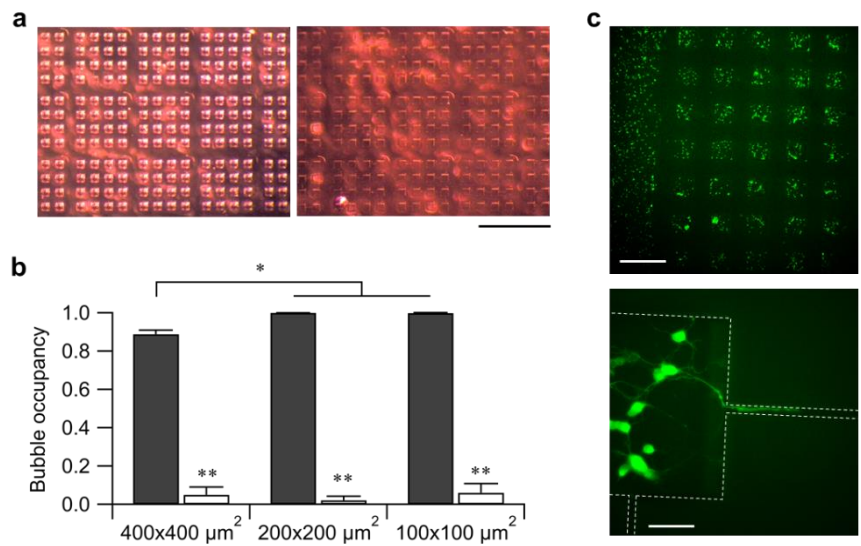

**Fig. 4 Neuronal culture.** (a) Gas entrapment in the reservoirs, which inhibits the plated cells from entering the device (left), can be substantially reduced using ethanol to mediate PDMS surface wetting (right). (b) Fraction of reservoirs occupied with air bubbles without (grey) and with (white) ethanol-mediated wetting. (c) Fluorescence micrograph of rat cortical neurons cultured in the 4-module micropattern for 3 days. Bird's-eye, low-magnification (top) and high-magnification (bottom) observations. The left side of the top panel is that of an area without the PDMS where neurons are randomly growing. In the bottom panel, the micropattern geometry is outlined with dashed lines to aid visualization. *$p < 0.05$, **$p < 0.01$ (two-sided Student's *t*-test). Scale bars, 1 mm (a), 500 μm (c, top), and 50 μm (c, bottom).

Fluorescent micrographs of the rat cortical neurons grown in 4-module micropatterns are shown in Fig. 4c. The neurons were stained with the neuronal marker NeuO to better visualize soma and neurites. The pre-plating procedure described above allowed the neurons to occupy almost all the reservoirs. High magnification observation of the microchannel area also showed a presumptive axon protruding from a soma and entering the microchannel (Fig. 4c).

**Network activity of the engineered mesoscale neuronal networks**

Finally, to validate the functionality of the micropatterned neuronal network grown in μFFs, rat cortical neurons were transfected with a calcium probe, GCaMP6s. Fluorescence intensity traces of spontaneous neural activity in a 4-module network with a line width of 10 μm is shown in Fig. 5a. The data is presented as the relative fluorescence intensity of neuron $i$, $f_i$ (= $[ΔF/F]_i$), which was calculated as follows:

$$f_i(t) = \left[\frac{\Delta F}{F}\right]_i = \frac{F_i(t) - F_{0(i)}}{F_{0(i)}},$$

where $F_i(t)$ is the fluorescence intensity of neuron $i$ at time $t$, and $F_{0(i)}$ is the background fluorescence intensity of neuron $i$.

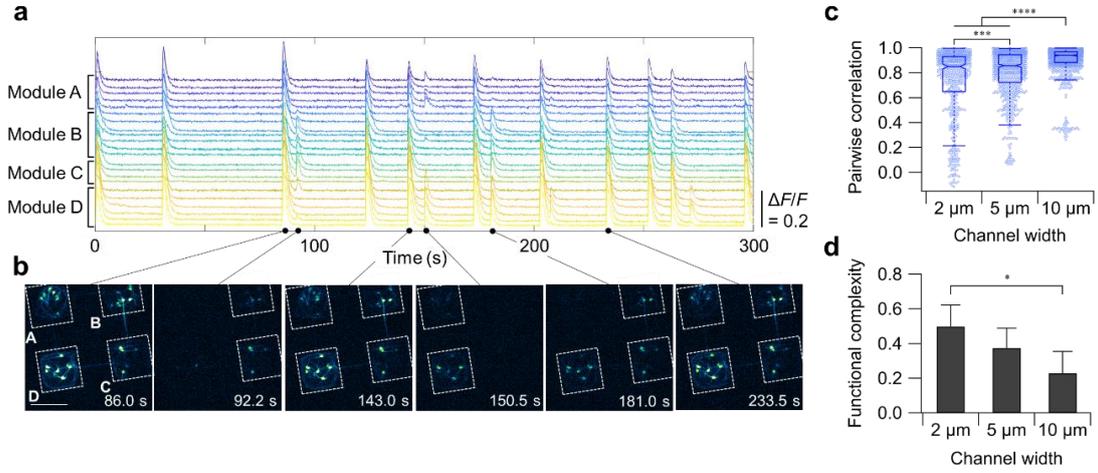

**Fig. 5 Functional analysis of spontaneous neural activity using fluorescence calcium imaging.** (a) Representative fluorescence traces from a 4-module network at 10 DIV. Spatial relationships of the four modules, A–D, are shown in (b). (b) Snapshots of the neuronal network at designated time points. The fluorescence micrographs were converted to ΔF/F images by calculating relative fluorescence intensity for each pixel. The modules activated at each time point are contoured with squares. Scale bar, 200 μm. (c) Distribution of pairwise correlation of spontaneous activity in neurons of the 4-module networks with the channel width of 2 μm ($n$ = 575), 5 μm ($n$ = 731), and 10 μm ($n$ = 942). 'Channel width' denoted in panels (c) and (d) refers to the line width of the photomask. The notch, box, and whisker represent the 95% confidence interval of the median, the interquartile range, and Tukey fences, respectively. Open circles are sample data points. ***, $p < 0.005$; ****, $p < 0.001$ (Mann-Whitney U-test). (d) Functional complexity for the 4-module networks with the channel width of 2 μm ($n$ = 5), 5 μm ($n$ = 7), and 10 μm ($n$ = 5). Data are shown as means ± S.D. *, $p < 0.05$ (two-tailed Student's t-test).

The spontaneous activity pattern of the prepared modular neuronal network was characterized by the coexistence of globally and locally synchronized activity. Globally synchronized activity, or network bursts,[24,26,27] was observed repetitively, e.g., at 86.0, 143.0, and 233.5 s (Fig. 5b). In addition to the typical activity of cultured neuronal networks, a rich variety of localized synchronizations were observed, e.g., at 92.2, 150.5, and 181.0 s (Fig. 5b). This observation reproduces previous experiments using microcontact-printed protein scaffolds, which showed that modular organization enriches the variety of spatiotemporal activity patterns exhibited within a single network.[24]

A comparison of 4-module networks with different channel widths revealed that the dynamical enrichment became more prominent when channel widths were further narrowed. As shown in Fig. 5c, the functional correlation between a pair of neurons became more broadly distributed between zero (desynchronized) and one (fully synchronous) in micropatterns with 2 and 5 μm channels. Here, a pairwise correlation $r_{ij}$ between neurons $i$-$j$ was calculated as $r_{ij} = [\langle f_i f_j \rangle - \langle f_i \rangle \langle f_j \rangle]/\sigma_{fi}\sigma_{fj}$, where $f_i = f_i(t)$ is the relative fluorescence intensity calculated here using time-dependent background intensity,[26] $\langle \cdot \rangle$ designate the time average, and $\sigma_x$ is the standard deviation of $x$. The broadening of the neural correlation resulted in the increase of functional complexity $\chi$, a statistical measure of integration-segregation balance in complex networks:[24,33]

$$\chi = 1 - \frac{1}{M} \sum_{\mu=1}^{m} \left| p_\mu(r_{ij}) - \frac{1}{m} \right|,$$

where $|\cdot|$ designate the absolute value, $m$ the number of bins (= 20), $M = 2(m-1)/m$ a normalization factor, and $p(\cdot)$ the probability. The value of $\chi$ has been shown to be maximized in the network topology obtained from *C. elegans*, cat, macaque, and human, and that any rewiring from the original topology decreases the value.[33] The mean functional complexity derived from the recordings of the 4-module networks are summarized in Fig. 5d, showing that a significantly higher balance of integration and segregation was achieved in networks comprised of the 2 μm-wide channel.

As discussed in the Introduction, plasma etching, capillary filling, manual air blowing, and gel spreading in open air have been previously proposed as methods to fabricate PDMS microstructures with well-defined through-holes.[19–23] After considering multiple approaches, we found that, for neuronal patterning applications, the open-air gel spreading approach is the simplest and most reliable way to produce μFFs. Although cell culture experiments using microfluidic devices with an array of sub-mm-sized reservoir holes have been reported,[12] this study is the first to report a fabrication procedure of such devices in detail. Under the increasing demand to study the structure-function relationships in biological neuronal networks at single-cell resolutions, the newly developed μFF technology provides a novel platform to engineer mesoscale neuronal networks.

## 4. Conclusions

We described herein a reliable protocol for producing μFFs which can be used to engineer network connectivity in mesoscale neuronal networks. The protocol consists of the standard fabrication of a master mould using SU-8, followed by a drop casting of a PDMS precursor gel. The thermally cured PDMS with a thickness of approximately 96 μm was sufficiently stable to be freestanding and be transferred to a coverslip pre-treated with poly-lysine. Gas entrapment in reservoir holes, which became an issue with decreased reservoir sizes, was efficiently suppressed by wetting the PDMS surface with ethanol and substituting the ethanol layer with a serum-containing medium. Finally, rat cortical neurons were cultured in a μFF with a modular structure consisting of four square-islands of 200 μm by 200 μm to examine whether the observations on microcontact-printed protein scaffolds could be replicated in the microfluidic device. Because of the high reproducibility, stability, and transferability of PDMS microfluidic devices, expanding the application of the microfluidic devices should impact researchers in the field of bioengineering and associated disciplines, including biology, medicine, and pharmacology.

**Conflicts of interest**

There are no conflicts of interest to declare.

**Acknowledgements**

We thank Dr. Hiroya Abe for fruitful discussion. This work was supported by the Japan Society for the Promotion of Science (Kakenhi Grant No. 18H03325) and by the Japan Science and Technology Agency (PRESTO: JPMJPR18MB and CREST: JPMJCR14F3, JPMJCR19K3). The experiments were partly carried out at the Laboratory for Nanoelectronics and Spintronics, Research Institute of Electrical Communication, Tohoku University.